\documentclass[%
 reprint,superscriptaddress,
 amsmath,amssymb,
 aps,
]{revtex4-2}

\usepackage{graphicx}
\usepackage{bm}
\usepackage{hyperref}

\usepackage{amsmath}
\usepackage{amssymb}
\usepackage{xcolor, soul}
\sethlcolor{magenta}
\usepackage{epstopdf}

\begin{document}

\preprint{APS/123-QED}

\title{Community attitudes towards the environmental cost of computational fluid dynamics research}

\author{Miranda van Heel}
\author{J. R. C. King}
\email{jack.king@manchester.ac.uk}
\affiliation{School of Engineering, the University of Manchester, UK
}%

\date{\today}

\begin{abstract}

Numerical simulations underpin much fluid dynamics research today. Such simulations often rely on large scale high performance computing (HPC) systems, and have a significant carbon footprint. Increasing the efficiency of data centers or the proportion of electricity coming from renewable sources can lessen the environmental impact of scientific computing to a degree, but the attitudes of researchers also play a role. There are many choices researchers make which influence the carbon footprint of simulations. To change behaviours around simulation use, it is first necessary to understand attitudes toward them. Here, we present a case study on fluid dynamics researchers based in the University of Manchester, UK. We find a low awareness of the carbon footprint of computations, compounded by a lack of knowledge of the specific hardware used to run simulations. There is a discrepancy between researchers self-declared attitudes towards reducing the carbon footprint of their work, and their actions and choices. Overall, researchers did not consider carbon footprint as important in their decision making, and we found no correlation between the impact and carbon cost of simulations. Improved education and awareness of the environmental impact of simulations is imperative in the interests of the sustainability of this field. 
    
\end{abstract}

\maketitle

\section{Introduction}\label{sec:intro}

Anthropogenic climate change, driven by greenhouse gas emissions, is one of the most important issues humanity faces in the modern age and it's effects are increasingly widespread. Worldwide, data centres and high performance computing (HPC) facilities are estimated to produce over 100 megatons of CO$_{2}$ per year, an amount is comparable to the total produced by American commercial aviation~\cite{GreenAlgorithms}. While the climate impact of aviation is widely known and discussed, the carbon footprint of HPC is much less visible yet rapidly increasing, particularly with the recent rise of artificial intelligence (AI) usage~\cite{AIrise}. Scientific computing is a major contributor to HPC usage, yet the research community's awareness of, and attitudes towards the carbon footprint of their work is unknown. 

Fluid dynamics is a £13.9 billion UK industry; of the approximately 45,000 people this industry employs, almost all use some aspect of computation in their work~\cite{UKFluid}. Many researchers in this field work towards solutions to the climate crisis, from wave structure interaction for offshore renewables~\cite{SamOceanEnvironment} to wake analysis of wind turbines~\cite{PabloTurbines}, thermal hydraulics for lifetime extension of nuclear reactors~\cite{HectorThermal} and  cleaner combustion technologies~\cite{CleanCombustion}. Even within these environmentally focussed fields, the impact of computational research is rarely reported.

Computational Fluid Dynamics (CFD), though a broad field spanning a range of resource intensities, is highly reliant on HPC systems. Many flows are highly multiscale, leading to simulations with a large number of degrees of freedom. Such large scale simulations using HPC facilities and data centres generate a significant carbon footprint~\cite{LowCarbonDatacenters}. Steps can be taken by data centres and HPC facilities to reduce their carbon footprint by increasing their efficiency, quantified by a Power Usage Efficiency (PUE), and updating their hardware (although production of new hardware generates a significant carbon footprint~\cite{CarbonHardware}), but as the demand for more computing power grows, perhaps something must be done to address simulation usage itself. Changing behaviours is one way to reduce the carbon footprint of simulations, and optimising simulation usage begins by understanding how researchers value their simulations.

To introduce more environmentally friendly practices, we first need to understand researchers' motivations and decision making processes. Why run a specific simulation? How important is the carbon footprint relative to the new science? Are they aware of the environmental impact? Do they consider alternative options to obtain the same insights more efficiently?

As a first step to answer these questions, and understand how researchers value their simulations, we surveyed and interviewed researchers working in the field of fluid dynamics at the University of Manchester. Here, HPC systems contribute approximately $50\%$ of overall IT electricity use~\cite{UoMsustainability}, whilst much of the remaining $50\%$ involves local workstations. Questions were designed to gauge opinions on the importance of simulations, determine what factors researchers are considering when running simulations, and evaluate attitudes towards greener practices. We asked participants for information on the simulations they run (hardware, number of cores/GPUs, wall-time, etc), and the carbon footprint of these was determined using the Green Algorithms calculator~\cite{GreenAlgorithms}. By comparing the carbon footprint of these simulations to other large sources of carbon emissions such as commercial flights, we hope to provide a relatable insight into the carbon footprint of the community's work.

\section{Findings}

\subsection{Demographic}

Researchers engaging with this project spanned career stages from post-graduate student to professor. With the majority of respondents sitting within the School of Engineering, more than half described their research as ``application focussed'', whilst more than half reported that high-performance computing was a particular focus of their research. A broad range of areas were covered, with respondents working across numerical methods (including Finite Volumes, Finite Differences, Finite \& Spectral Element Methods, pseudo-spectral methods, Lattice Boltzmann methods, Smoothed Particle Hydrodynamics and other mesh-free methods), and modelling techniques (with Direct Numerical Simulations most frequently reported, and a significant proportion conducting Large Eddy Simulations, whilst only one respondent reported using Reynolds Averaged Navier-Stokes (RANS) formulations). The most frequently reported research areas were Fluid-Structure interactions, heat transfer and turbulence, and offshore and Renewable energy, but respondents also reported work in combustion, multi-phase and interfacial flows, aero-acoustics, environmental flows and rheology. Most respondents, $91\%$, reported primarily using open source codes. $64\%$ used in-house codes, and only $18\%$ reported using commercial codes. Approximately $100$ researchers were invited to participate, and $24$ responded to the survey; a relatively low uptake, perhaps indicative of the relative lack of importance with which the topic is viewed in the field. 

\subsection{Carbon footprint}
All respondents agreed that they considered simulation results imperative for advancing understanding in their respective research fields. Despite this, respondents awareness of the carbon footprint of their computations was mixed, with equal numbers aware and unaware. Estimates of carbon footprint were made using the Green Algorithms calculator, which combines information about hardware, run time and data center location to estimate emissions. Specific PUEs of each data center were used where possible, and otherwise the worldwide average of $1.67$ was used~\cite{GreenAlgorithms}. CPU and GPU usage factors were assumed to be $100\%$, providing a conservative estimate of the computing cost of simulations. We note that whilst the UK Tier 1 HPC system (ARCHER2) is exclusively powered by green electricity, we have calculated the carbon footprint of simulations assuming the UK average electricity mix: if a simulation using electricity from renewables were not run, that green electricity could be utilised elsewhere, reducing demand on the national grid. In our opinion, a system running on renewable energy should not provide carte blanche to disregard the power consumed by simulations on that system. When looking at a sample of approximately $40$ papers in the field citing use of University of Manchester HPC systems over the past $3$ years, only $8$ contained enough information (hardware, core-counts, walltime) to estimate the carbon footprint of any simulation used. Either researchers are unaware of the hardware used in their computations or do not consider it important enough to include in publications. We asked researchers to estimate their typical HPC monthly usage (in CPU- or GPU-hours). $35\%$ were unable to provide an estimate, with one responding ``have no clue''. Of those who provided an estimate, this varied from $10$ GPUh/month to $10^{6}$ CPUh/month, highlighting the range in levels of HPC use within the field. This lack of awareness highlights the challenge in attitudes: if many researchers do not even roughly track their simulation usage, how can they have knowledge of their potentially significant carbon footprint, even before considering ways to reduce it? The lack of awareness of the carbon cost of computational research was further highlighted in interviews, where participants were universally unable to provide estimates of their research carbon footprint, and none were aware of any tools to help with such calculations. Further, many academics admitted that estimating the carbon footprint of their research was something that they should have done previously.

Researchers were asked for relevant information on a large-scale simulation they had completed recently. The classification of a simulation as ``large-scale'' was deliberately left to interpretation, and responses ranged from simulations on a single GPU to those using thousands of CPUs on Tier 1 (national) HPC facilities. Once again, many were unable to give information on the hardware they used, with only $33\%$ able to identify the specific hardware used. This is troubling as the thermal design power (TDP) of a CPU or GPU can significantly vary, and is a main factor in determining the carbon footprint of a simulation~\cite{GreenAlgorithms}. The carbon footprint of the researchers' simulations varied significantly due to the interpretation of ``large scale'' simulation, with the smallest only generating $1.43$kgCO$_2$e, and the largest generating $56\times{10}^{3}$ kgCO$_{2}$e, the equivalent of 24 flights from New York to Melbourne \cite{GreenAlgorithms}. The mean simulation size, in terms of emissions, was $9134$ kgCO$_2$e, equivalent to $3.95$ flights from New York to Melbourne. $31\%$ of participants thought that they could have achieved the results of their simulation at lower computational cost, and when asked how, the responses varied from using different hardware such as comparing using a GPU to multiple CPU cores to significantly speed up the simulation, or using symmetric boundary conditions to cut the cost in half. Further, researchers reported that of the typical simulations they run, only $56\%$ actually give useful results and $71\%$ of researchers disagreed that they only run simulations when absolutely necessary, indicating that participants are aware that they potentially overuse computational resources. 

\subsection{Decision making}

There are many factors that influence decision making when selecting certain numerical methods and techniques, e.g. accuracy/precision, speed/costs, and scalability. Participants scored (on a scale from $0$ to $1$) a range of factors based on how important they considered them. Accuracy, precision and convergence were ranked as most important (mean score $0.85$) by participants, closely followed by measures relating to scalability, efficiency and speed of simulations (mean score $0.78$). The carbon footprint was ranked as least important (mean score $0.37$). These factors are not necessarily independent. A highly accurate and efficient numerical method should yield results at a lower carbon cost than an inaccurate and inefficient one, and $94\%$ of respondents agreed that they place importance on computational efficiency in their work. However, the low score attributed to the carbon footprint highlights that it is a factor people are not directly considering. Researchers may be seeking to use methods which provide them with accurate solutions cheaply, but if this results in a low carbon footprint, that is perhaps a secondary benefit. Of all the factors assessed, the importance given to carbon footprint saw the greatest variation across participants (standard deviation $\sigma=0.30$). In contrast, precision accuracy and convergence ranked highly across all respondents ($\sigma=0.10$), as did measures relating to scalability, speed and efficiency ($\sigma=0.14$). Consideration of the carbon footprint of a simulation was not found to be correlated with the carbon footprint of simulations, and researchers self-reported typical HPC use. 

Overall $69\%$ of respondents agreed that they conducted research in an environmentally friendly manner, and $75\%$ disagreed that they could do more to reduce the carbon footprint of their research, although $63\%$ acknowledged they do not only run simulations when absolutely necessary. In response to the statement ``My research is conducted in an environmentally friendly manner'', only $12\%$ strongly agreed or strongly disagreed. This apparent uncertainty was reinforced in interviews: on the environmental cost of simulations, one academic acknowledged they were ``not aware of the scale of it''. When considering a specific large scale simulation, only $22\%$ felt they could have obtained the same insight at cheaper computational cost, and of the $39\%$ who experienced issues with their simulation, only $29\%$ said further testing could have avoided these issues. 

\subsection{Impact}

The majority of respondents agreed that they regularly evaluated the impact of their research on society. A topic that was raised in multiple interviews was the juxtaposition between ``big picture'' research goals in support of a low carbon future, and the carbon footprint of the research itself. One academic with research interests in offshore renewables said: ``you'd like to think you have a net positive affect and overall I think that's the case, but not necessarily for individual projects.'' We note that such a conflict equally arises in conference travel in these fields; researchers may ask whether the net benefit of their individual attendance at a meeting outweighs the carbon cost of attending. Indeed, the individual carbon footprint of climate research has been shown to affect how their advice is perceived~\cite{attari_2016}. Whether such perceptions apply in the context of the carbon footprint of computational research is unknown.

\begin{figure}[h]
    \centering
    \includegraphics[width=\columnwidth]{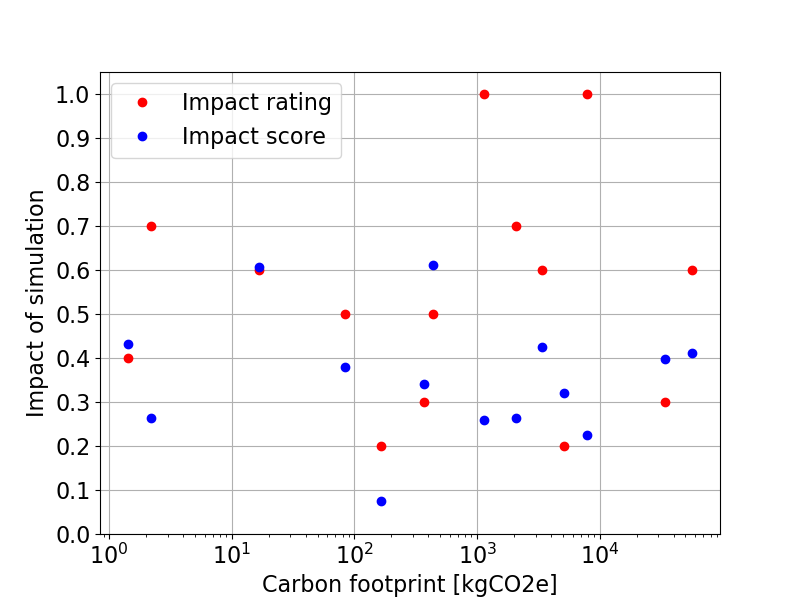}
    \caption{The impact of a large scale simulation as rated by survey respondents (impact rating) and as calculated from simulation information (impact score), compared to the carbon footprint of that simulation.}
    \label{fig:impactscores}
\end{figure}

Research with a larger carbon footprint may be justified by a significant impact. When describing a recent large scale simulation, respondents were asked to rate its impact. To avoid bias towards user opinion, this impact rating was compared to an impact score based on other information about their simulations such as the ways in which it contributed to their field and how far they expected it to reach, from individual data analysis to a main part of a publication. Both measures are plotted in Figure~\ref{fig:impactscores}, with the self-reported rating in red, and the calculated score in blue, against the estimated carbon footprint of the calculation. There is no clear correlation between the impact rating given by respondents and the calculated impact score. Impact is an ambiguous concept, and it follows that it is difficult to objectively quantify the impact of a given simulation. However, there is no correlation between either measure of impact and carbon footprint. Although this is a small sample, there is no evidence that larger simulations are leading to larger impact. 

Often issues are encountered when running a simulation, such as a failure for the numerical method to converge or behave as expected, or the discovery during subsequent analysis that incorrect parameters have been chosen (e.g. discovering numerical artefacts indicative of an insufficient resolution). When running large scale simulations, it is standard practice to run smaller test simulations to avoid wasting computational resources and all respondents confirmed that they did this. Although $39\%$ reported they still encountered such issues in their large scale simulation, all participants agreed that they had done as much as possible in advance to ensure large scale simulations ran correctly, despite reporting that a significant $40\%$ of typical simulations they run go wrong and have to be repeated. Respondents reported on average that $56\%$ of simulations run yielded ``useful results'', that only $29\%$ of simulations run yield results directly used in publications, and that $38\%$ ``go wrong'' and have to be repeated with adjustments. 

These findings suggest there may be a dichotomy between our ideals and values around the environmental impact of our research in theory, and our behaviours in practice, but we also see an opportunity for improvement. Perhaps a lack of accountability (or even awareness of the carbon footprint) allows us to turn a blind eye to the environmental cost. The lack of consideration of the carbon cost of computations was a clear theme we observed throughout our interactions with the community. If it were easier to calculate the carbon footprint, or if an HPC system provided an estimate of the carbon footprint of a simulation by default, perhaps it would be harder for us to ignore this cost. If attitudes changed and reporting the carbon footprint in publications became the norm, how would our decision making process differ?

\subsection{Attitudes}
There are many reasons we run simulations, and the motivation for specific cases varied widely across respondents, from validation of numerical methods, to obtaining fundamental insight into physical systems, to confirming hypotheses suggested by experimental results. One simply reported ``it was essential'', perhaps unwittingly highlighting the issue we face. Objectively, it is hard to see how a computational fluid dynamics simulation is ever \emph{essential}, but for an individual researcher, conducting a simulation \emph{feels} essential; what are the consequences of not running the simulation? Perhaps without it a paper cannot be finished. Or in second guessing referrees comments, we might see it as necessary to get a paper through peer review. We have heard numerous anecdotes from academics of pressure from referress for further large scale simulations for negligible benefit. Perhaps it is necessary in order to complete a PhD. Perhaps a paper based on this simulation will form the basis for a future grant application. 

These are all legitimate reasons, and we are not unsympathetic to such arguments; within the confines of available resources, academic freedom rightly allows the individual to decide how to proceed. However, would the outcome of these deliberations differ if greater consideration and emphasis were placed on the carbon cost? 

\section{Recommendations}

Computational Fluid Dynamics is a field of research with a large carbon footprint. The carbon cost of simulations is generally not reported, and researchers appear to have a low awareness of the environmental impact of their work. To ensure our field of research remains sustainable in the future, it is imperative that this changes. We do not propose a restriction on academic freedoms in this arena, and we acknowledge that there are other aspects of our field (e.g. HPC hardware/infrastructure) which influences the carbon footprint of our research, and over which we have less control. 

Education is key. Researchers cannot make informed decisions if they are unaware of the carbon cost of their simulations. In the UK, many public sector organisations have sustainability goals. Increased awareness of the environmental impact of simulations is directly aligned with such goals. Training should be provided to ensure those running large scale simulations are better informed of the carbon footprint; that they understand how different choices (e.g. which numerical method) impact this; and give it increased consideration when making decisions.

A greater awareness of these issues has the potential to drive a change in attitudes. Those running HPC systems have a role to play: if HPC systems provide an estimate of the carbon footprint of a simulation by default, researchers would be better informed and more aware of the cost, and we believe this should be the norm. We recommend inclusion of the carbon footprint of simulations in publications. If this practice becomes widespread, then changes in community attitudes will follow, with an increased acceptance that environmental arguments have a place in the decision making process around computationally intensive research.    

\section*{Acknowledgements}

This work was funded by the Royal Society, via a University Research Fellowship (URF\textbackslash R1\textbackslash 221290). We are grateful to Lo\"{i}c Lannelogue for creating the Green Algorithms calculator, and for support in its use. We are grateful for assistance given by Research IT at the University of Manchester.  

\bibliography{references}

\end{document}